\documentclass{article}

\usepackage{appendix}

\usepackage[margin=1in]{geometry}
\usepackage{setspace}
\usepackage[utf8]{inputenc}
\usepackage[T1]{fontenc}
\usepackage{lmodern} % Better font rendering

\usepackage{algorithm}
\usepackage{algpseudocode}
\usepackage{amsmath, amssymb, amsthm, amsfonts, mathtools}
\usepackage{bm} % bold math symbols

\usepackage{graphicx}
\usepackage{caption}
\usepackage{subcaption}
\usepackage{booktabs} % Better table formatting

\usepackage[backend=biber, style=ieee, sorting=none]{biblatex} % or apa,
\usepackage[hidelinks]{hyperref}
\hypersetup{
    breaklinks=true,
    colorlinks=False
}

\newtheorem{theorem}{Theorem}
\newtheorem{lemma}[theorem]{Lemma}
\newtheorem{definition}{Definition}

\title{A Probabilistic Computing Approach to the Closest Vector Problem for Lattice-Based Factoring}

\author{
    Max O. Al-Hasso\textsuperscript{1,*}
    \and Marko von der Leyen\textsuperscript{1,2}
}

\date{}

\begin{document}

\maketitle
\begin{center}
    \textsuperscript{1} Quantum Dice Ltd, 264 Banbury Road, Oxford, OX2 7DY, United Kingdom\\
    \textsuperscript{2} Department of Physics, University of Oxford, Oxford, OX1 3PU, United Kingdom
    \\
    \vspace{0.25cm}
    \textsuperscript{*}\texttt{max.al-hasso@quantum-dice.com}
\end{center}

\begin{abstract}

The closest vector problem (CVP) is a fundamental optimization problem in lattice-based cryptography and its conjectured hardness underpins the security of lattice-based cryptosystems. Furthermore, Schnorr's lattice-based factoring algorithm reduces integer factoring (the foundation of current cryptosystems, including RSA) to the CVP. Recent work has investigated the inclusion of a heuristic CVP approximation `refinement' step in the lattice-based factoring algorithm, using quantum variational algorithms to perform the heuristic optimization. This coincides with the emergence of probabilistic computing as a hardware accelerator for randomized algorithms including tasks in combinatorial optimization. In this work we investigate the application of probabilistic computing to the heuristic optimization task of CVP approximation refinement in lattice-based factoring. We present the design of a probabilistic computing algorithm for this task, a discussion of `prime lattice' parameters, and experimental results showing the efficacy of probabilistic computing for solving the CVP as well as its efficacy as a subroutine for lattice-based factoring. The main results found that (a) this approach is capable of finding the maximal available CVP approximation refinement in time linear in problem size and (b) probabilistic computing used in conjunction with the lattice parameters presented can find the composite prime factors of a semiprime number using up to 100x fewer lattice instances than similar quantum and classical methods.

\end{abstract}

\section{Introduction}

% Probabilistic computing is an emerging domain-specific architecture for accelerating probabilistic algorithms and applications utilizing inherently stochastic hardware. 
% The basic unit of information in these systems is the probabilistic bit (p-bit) which fluctuates between 0 and 1 with a tuneable bias.

% Probabilistic computing is an emerging domain-specific architecture which utilizes inherently stochastic hardware for accelerating probabilistic algorithms and applications. 
% Inspired by The basic unit of information in these systems is the probabilistic bit, 

% Probabilistic computing is an emerging domain-specific architecture for accelerating probabilistic algorithms with much recent research interest focused on its application in combinatorial optimization/

% Probabilistic computing is an emerging computer architecture designed to accelerate probabilistic algorithms, with recent research highlighting its potential in solving hard combinatorial optimization problems.

Probabilistic computing is an emerging computational architecture that is currently receiving increased research attention due to its potential to accelerate randomized algorithms including tasks in combinatorial optimization. Inspired by physical stochastic processes, theses systems are comprised of a network of interconnected probabilistic bits (p-bits) that fluctuate between two states (0,1) according to a controlled probability calculated from the states of neighboring p-bits. P-bits can be realized in hardware using CMOS+  where conventional CMOS hardware is used in conjunction with a hardware entropy source such as magnetic nanodevices \cite{chowdhuryFullStackViewProbabilistic2023, singhCMOSStochasticNanomagnets2024} and, more recently, quantum random number generators \cite{shelbayaSelfcorrectingHighSpeedPhotonic}. Probabilistic computing shows promise as a hardware solver for combinatorial optimization problems \cite{aaditComputingInvertibleLogic2021, aaditMassivelyParallelProbabilistic2022, searleVirtuallyConnectedProbabilistic, hasselgrenProbabilisticComputingOptimization} and as a contender to methods such as simulated annealing \cite{vanlaarhovenSimulatedAnnealingTheory1987}, quantum annealing \cite{kadowakiQuantumAnnealingTransverse1998}, and variational quantum algorithms such as the quantum approximate optimization algorithm (QAOA) \cite{farhiQuantumApproximateOptimization2014}. In this paper, we present a method for applying probabilistic computing to combinatorial search problems relevant to cryptanalysis.

Integer factorization is central to modern cryptography due to the conjectured hardness of the problem. It is the foundation of many widely used cryptosystems including RSA \cite{rivestMethodObtainingDigital1978}. While Shor's algorithm \cite{shorPolynomialTimeAlgorithmsPrime1997} is capable of finding prime factors in polynomial time given a large, fault-tolerant quantum computer it is limited by present-day quantum hardware. Among classical factoring methods, recent work \cite{yanFactoringIntegersSublinear2022a, priestleyPracticallyScalableApproach2025, tesoroQuantumInspiredFactorization2024c, aboumradQuantumClassicalCombinatorial2023, khattarCommentFactoringIntegers2023} has focused on Schnorr's lattice-based factoring algorithm \cite{schnorrFactoringIntegersComputing1991, schnorrFastFactoringIntegers2021}, which reduces integer factoring to finding solutions to many instances of the closest vector problem (CVP). This reduction highlights a promising intersection: the CVP is both central to lattice-based factoring and potentially well suited to hardware-accelerated probabilistic computing. Beyond integer factoring, the CVP and the related shortest vector problem are core problems in lattice-based cryptography - a leading candidate for post-quantum cryptography (PQC) \cite{computersecuritydivisionPostQuantumCryptographyStandardization2017}. Hardware solvers for the CVP and other lattice problems will be of increasing importance in the PQC era.

In this work, we investigate the application of probabilistic computing to solving the CVP in the context of lattice-based factoring. Our contributions include:

\begin{itemize}
    \item An analysis of `prime lattice' parameters supported by experimental results.
    \item A mapping from the CVP to a p-bit network and an evaluation of probabilistic computing as a combinatorial solver for the CVP.
    \item An experimental evaluation of probabilistic computing as a method for lattice-based sieving (i.e. as a subroutine in lattice-based factoring) and comparison to QAOA \cite{yanFactoringIntegersSublinear2022a, aboumradQuantumClassicalCombinatorial2023}.
\end{itemize}

Importantly, we note that the best classical algorithm for factoring cryptographically relevant compound integers is the general number field sieve (GNFS) \cite{lenstraDevelopmentNumberField1993, briggsIntroductionGeneralNumber1998}, with the largest known successful RSA number factorization utilizing this method \cite{zimmermannFactorizationRSA2502020}. This work is concerned with solving the CVP and improvements to lattice-based factoring, making no claims of speed up with respect to GNFS.

\section{Background}
\subsection{Probabilistic computing}
% Probabilistic computing arose from the same notion of `natural computing' as quantum computing, wherein a physical system with some desirable property is leveraged to perform computation via a natural mapping from device to algorithm. In quantum computation 

Probabilistic computing, like quantum computing, arose from the idea of `natural computing' wherein physical systems with advantageous properties are used to perform computation through natural system-algorithm mappings \cite{feynmanSimulatingPhysicsComputers1982, chowdhuryFullStackViewProbabilistic2023}. As such, the p-bit is typically described as an intermediate between the classical bit and the qubit. Where the bit stores a fixed state (either 0 or 1) and the qubit exists in a superposition (both 0 and 1), the p-bit is a random variable with a parameterized Bernoulli distribution, and the state of the p-bit is sampled from this distribution each time it is observed (either 0 or 1 with some controlled probability). In \cite{camsariStochasticBitsInvertible2017} the p-bit is defined as a binary stochastic neuron realized in hardware. 
%g-bits?

In the context of combinatorial optimization, a probabilistic computer is typically configured to simulate an Ising model, where the optimization objective corresponds to the ground state of the system. In this configuration, for a given p-bit $i$, the bias $b_i$ is calculated from the state of neighboring p-bits according to:

\begin{equation} \label{eq:ising-bias}
    b_i = \sum_j J_{ij}s_j + h_i 
\end{equation}

Where $J$ is a symmetric matrix describing the bidirectional interactions between any two given p-bits, $h$ is a constant applied to each p-bit, and $s_j$ is the sampled state of p-bit $j$.

A recent work \cite{jungQuantuminspiredProbabilisticPrime2023} introduced the concept of the virtually connected Boltzmann machine (VCBM), a model of probabilistic computing where the bias for a given p-bit is computed directly (shown below in Definition \ref{def:pbit-bias}). This approach is more expressive as it is not limited to two-body interactions as in the Ising model and furthermore does not require precomputation of the $J$ and $h$ matrices. However, the computational cost of an arbitrary bias calculation is likely to be greater than the relatively low cost of Equation \eqref{eq:ising-bias}. In this work we choose to use the VCBM approach, but as the energy function (Equation \eqref{eq:vcbm-energy-func}) used is quadratic there exists a straightforward mapping between the two approaches.

% - formal p-bit definition
% - feynman?
% - g-bits
% - p-bit abstraction, p-bit is a bernouli distribution

\subsection{Integer factoring}
Classical algorithms for factoring semiprimes, including the quadratic field sieve \cite{pomeranceSmoothNumbersQuadratic2008}, Schnorr's algorithm \cite{schnorrFactoringIntegersComputing1991, schnorrFastFactoringIntegers2021}, and the best-in-class General Number Field Sieve \cite{lenstraDevelopmentNumberField1993, briggsIntroductionGeneralNumber1998}, do so via the congruence of two squares method (Lemma \ref{lemma:congruence-of-squares}). As such, these algorithms share a similar two-phase structure. In the first phase, the algorithm searches for and collects a set of $B$-smooth numbers or pairs of $B$-smooth numbers of which a subset can be used to generate a congruence of squares. In the second phase, linear algebra is used to find the subset of collected relations that yields a congruence. If the congruence that is generated is trivial, additional $B$-smooth numbers must be found and the linear algebra phase repeated. 
%In the case of Schnorr's algorithm, this is done according to Definition . 

Recently, lattice-based factoring method has received increased attention due to a proposal by Yan et al. \cite{yanFactoringIntegersSublinear2022a} to include a subroutine which uses QAOA to refine the CVP solution. The broader claims of this paper have been thoroughly disputed \cite{aboumradQuantumClassicalCombinatorial2023, khattarCommentFactoringIntegers2023, priestleyPracticallyScalableApproach2025, grebnevPitfallsSublinearQAOAbased2023}, yet the CVP refinement subprocess remains an interesting extension to the original lattice-based factoring algorithm and to the CVP more generally. Subsequent work has investigated improvements to QAOA \cite{priestleyPracticallyScalableApproach2025} and the application of tensor network methods to the refinement step \cite{tesoroQuantumInspiredFactorization2024c}.

\section{Preliminaries}
\subsection{Integer factoring}

\begin{definition} \textbf{Smooth Number.} \cite{pomeranceSmoothNumbersQuadratic2008}

An integer $x \in \mathbb{N}$ is $B$-smooth if all the prime factors of $x$ are less than or equal to $B$. 
\end{definition}

The notation $\pi(B)$ is used to mean the number of primes in the interval $[1,B]$ and $B$ is referred to as the smoothness bound. Furthermore, the set of primes less than or equal to $B$ is referred to as the factoring basis.

\begin{definition} \textbf{Exponent Vector.} \cite{pomeranceSmoothNumbersQuadratic2008}

The exponent vector of a $B$-smooth integer $x \in \mathbb{N}$ is the vector $\mathbf{e}\in \mathbb{N}^{\pi(B)}$ such that
\begin{equation}
    x=\prod_{i=1}^{\pi(B)} p_i^{e_i}
\end{equation}
where $p_i$ is the $i$-th prime. 
\end{definition}
This concept can be extended to represent negative integers by prepending $e_0\in \{0,1\}$ to $\mathbf{e}$ and including $p_0=-1$ in the factoring basis.

\begin{definition} \textbf{Smooth Relation Pair (sr-pair, fac-relations)} \cite{schnorrFactoringIntegersComputing1991}

A pair of numbers $(u,v)$ is a $B$-smooth relation pair if both $u$ and $v$ are $B$-smooth and $u-vN$ is also $B$-smooth, for a given semiprime $N$. 
\end{definition}
% Smooth relation pairs are also referred to as sr-pairs or fac-relations.

\begin{lemma} \label{lemma:congruence-of-squares} \textbf{Nontrivial factor from Nontrivial Congruence of Squares.} \cite{pomeranceSmoothNumbersQuadratic2008}

Let $N\in \mathbb{N}$ be a semiprime integer and $x, y \in \mathbb{Z}$ such that
\begin{equation}
    x^2 \equiv y^2 \mod N \quad \text{and} \quad    x \not\equiv \pm y \mod N
\end{equation}

Then, 
\begin{equation}
    \gcd(x-y, N) \quad \text{and} \quad \gcd(x+y, N)
\end{equation}
are non-trivial factors of $N$, where $\gcd$ denotes the greatest common divisor function.
\end{lemma}

\begin{definition} \label{def:congruence-of-squares}\textbf{Finding a Congruence of Squares from Smooth Relation Pairs.} \cite{schnorrFactoringIntegersComputing1991, priestleyPracticallyScalableApproach2025}

For a given semiprime $N$ and smoothness bound $B=p_n$ where $p_n$ is the $n$-th prime, an sr-pair may be written in the form
\begin{equation}
    u=\prod_{i=0}^n p_i^{e_i}, \quad u-vN=\prod_{i=0}^n p^{e'_i}_i
\end{equation}

By construction, we have
\begin{equation}
    u\equiv u-vN \mod N \implies \frac{u-vN}{u}\equiv1 \mod N
\end{equation}

Furthermore,
\begin{equation}
    \frac{u-vN}{u} = \frac{\prod_{i=0} p^{e'_i}_i}{\prod_{i=0} p^{e_i}_i}=\prod_{i=0}^n p_i^{(e'_i-e_i)}:=\prod_{i=0}^n p_i^{\tilde{e_i}}
\end{equation}

Where the exponent vector $\mathbf{\tilde{e}} = (\tilde{e_0}, \ldots, \tilde{e_n})$ denotes the sr-pair ratio. Given a set of $n+2$ sr-pairs corresponding to a semiprime $N$, and therefore $n+2$ sr-pair ratio exponent vectors $\{\mathbf{\tilde{e}}_i\}_{i=0}^{n}$, we can find a congruence of squares by finding a solution $\mathbf{\tau}\in\{0,1\}^{n+2}$ to the system of linear equations
\begin{equation}
    \sum_{j=0}^{n+1} \tau_j \tilde{e}_{i,j}\equiv0\mod 2, \quad \text{for } i=0,\ldots, n
\end{equation}

% This is equivalent to finding the nullspace (or kernel) of the matrix $D\in \mathbb{F}_2^{(n+2)\times(n+1)}$ over the Galois field with $2$ elements, with $D$ defined as
% \begin{equation}\label{eq:system-of-linear-equations}
%     D :=
%     \begin{bmatrix}
%     \tilde{e}_{0,0} & \tilde{e}_{0,1} & \cdots & \tilde{e}_{0,n} \\
%     \tilde{e}_{1,0} & \tilde{e}_{1,1} & \cdots & \tilde{e}_{1,n} \\
%     \vdots & \vdots & \ddots & \vdots \\
%     \tilde{e}_{n,0} & \tilde{e}_{n,1} & \cdots & \tilde{e}_{n,n} \\
%     \tilde{e}_{n+1,0} & \tilde{e}_{n+1,1} & \cdots & \tilde{e}_{n+1,n}
%     \end{bmatrix} \mod 2
% \end{equation}

% The nullspace is therefore,
% \begin{equation}\ker(D)=
%     \{ \mathbf{\tau} \in \mathbb{F}_2^{n+2} | D \cdot \mathbf{\tau} = 0 \}
% \end{equation}
% Which is the set of solutions to the system of linear equations \eqref{eq:system-of-linear-equations}. 

A congruence of squares can be generated by 
% \begin{equation}
%     X = \prod_{\tilde{e}_{i,j}>0} p_i^{ \frac{1}{2}\sum_{j=1}^{n+1} \tau_j \tilde{e}_{i,j}} \mod N, \quad Y = \prod_{\tilde{e}_{i,j}<0} p_i^{ -\frac{1}{2}\sum_{j=1}^{n+1} \tau_j \tilde{e}_{i,j}} \mod N
% \end{equation}

\begin{equation}
    X:= \prod_{i=0}^n p_i^{\frac{1}{2}\sum_{j=0}^{n+1}\tau_j(e_{i,j}+e'_{i,j})} \mod N, \quad Y := \prod_{i=0}^n p_i^{\sum_{j=0}^{n+1}\tau_je'_{i,j}} \mod N
\end{equation}

By Lemma \ref{lemma:congruence-of-squares}, if $X\not\equiv \pm Y \mod N$, this construction yields two nontrivial factors by computing $\gcd(X+Y, N)$ and $\gcd(|X-Y|, N)$. However, if $X\equiv\pm Y\mod N$ another $\tau$ must be used. If there is no $\tau$ such that $X\not\equiv \pm Y \mod N$, additional sr-pairs must be collected, a larger system of linear equations constructed, and the new $\tau$ trialed. 
\end{definition}

% Schnorr's algorithm is split into two phases. In the first phase, the goal is to find at least $n+2$ sr-pairs. This is necessary to guarantee the nullspace contains at least one non-trivial (non-zero vector) solution. 

% I think i was incorrect for this bit \/
% That is, in equation (ref) there are $n+2$ vectors in $\mathbb{F}_2^{n+1}$ implying at least one  implying $\operatorname{rank}(D) < n+1$. Therefore, by the rank-nullity theorem 
% $$\operatorname{nullity}(D) = \dim(D) - \operatorname{rank}(D) =n+1-\operatorname{rank}(D)>0$$

\subsection{Probabilistic computing}
\begin{definition}\label{def:p-bit} \textbf{Probabilistic Bit (p-bit).}

A probabilistic bit is a unit of information that fluctuates between the states $\{0,1\}$ according to a tuneable bias. For a given bias $b\in\mathbb{R}$, the probability that the instantaneous state of a p-bit is $1$ is given by the logistic function
\begin{equation}
    P(s=1|b)=\frac{1}{1+\exp(-b)}
\end{equation}
Similarly, 
\begin{equation}
    P(s=0|b)=1-\frac{1}{1+\exp(-b)}=\frac{1}{1+\exp(b)}
\end{equation}

\end{definition}

\begin{definition} \label{def:pbit-bias} \textbf{Calculating p-bit bias.}

Given the collective state of an $n$ p-bit system $\mathbf{s}\in\{0,1\}^n$, the index of p-bit $i\in[1,n]$, and an energy function $E:\{0,1\}^n\rightarrow\mathbb{R}$, the bias of the $i$-th p-bit is given by
\begin{equation}
    b_i=E(\mathbf{s}|s_i=0)-E(\mathbf{s}|s_i=1)
\end{equation}

where $E(\mathbf{s}|s_i=0)$ and $E(\mathbf{s}|s_i=1)$ denote the energy of the state $\mathbf{s}$ with the value of the $i$-th p-bit set to $0$ and $1$ respectively. 
\end{definition}

\subsection{Lattices}

\begin{definition} \textbf{Lattice.}

A lattice is generated by the linear combinations of a set of linearly independent vectors. That is, given a set of $n$ linearly independent vectors in $m$-dimensional space, $\mathbf{v}_1,\ldots, \mathbf{v}_n\in \mathbb{R}^m$

\begin{equation}
    \mathcal{L}(\mathbf{v}_1, \ldots, \mathbf{v}_n) = \{a_1\mathbf{v}_1+\ldots+ a_n\mathbf{v}_n \,|\, a_1, \ldots, a_n \in\mathbb{Z}\}
\end{equation}

\end{definition}

\begin{definition} \textbf{Basis.}

    A basis of a lattice $\mathcal{L}$ is a set of linearly independent vectors $\mathbf{b}_1, \ldots, \mathbf{b}_n\in\mathbb{R}^m$ which generate $\mathcal{L}$. $\mathbf{B}=[\mathbf{b}_1,\ldots\mathbf{b}_n]\in\mathbb{R}^{(m\times n)}$ is the basis matrix of $\mathcal{L}(\mathbf{B})$ such that
    \begin{equation}
        \mathcal{L}(\mathbf{B})=\{\mathbf{Bx}|\mathbf{x}\in\mathbb{Z}^n\}
    \end{equation}
\end{definition}

\subsubsection{Closest vector problem}
\begin{definition} \textbf{Distance Operator.} \cite{stephens-davidowitzCVPBabaisAlgorithm2016}
    
    Given a target vector $\mathbf{t} \in \mathbb{R}^n$ and a lattice $\mathcal{L}\subset \mathbb{R}^n$, we define 
    \begin{equation}
        \operatorname{dist}(\mathcal{L}, \mathbf{t}):=\min_{x\in \mathcal{L}} ||x-\mathbf{t}||
    \end{equation}

\end{definition}

\begin{definition}\label{def:gap-cvp} $\boldsymbol{\gamma}$\textbf{-CVP} \cite{stephens-davidowitzCVPBabaisAlgorithm2016}

    For any approximation factor $\gamma=\gamma(n)\geq 1$, the $\gamma$-Closest Vector Problem ($\gamma$-CVP) is defined as follows. The input is a bias for the lattice $\mathcal{L}\subset \mathbb{R}^n$ and a target vector $\mathbf{t}\in \mathbb{R}^n$. The goal is to output a lattice vector $\mathbf{x}\in\mathcal{L}$ with $||\mathbf{x}-\mathbf{t}||\leq \gamma\cdot \operatorname{dist}(\mathbf{t}, \mathcal{L})$.
\end{definition}

% \begin{definition}\label{def:sr-pairs-from-lattice-points} Constructing sr-pairs from lattice points \cite{schnorrFactoringIntegersComputing1991, tesoroQuantumInspiredFactorization2024c}

% Given a lattice point $\mathbf{b}$ in a prime lattice  
    
% \end{definition}

\section{Problem Formulation}\label{sec:problem_formulation}

The collection phase of semiprime factoring, can be reduced to solving many instances closest vector problem as first described by Schnorr in \cite{schnorrFactoringIntegersComputing1991}. By constructing so-called \textit{prime lattices} and generating a CVP target dependent on the semiprime $N$ it is conjectured is that lattice points close to the target have a high likelihood of generating an sr-pair with respect to $N$.

%via the process described in definition (ref to sr-pair from lattice point definition).

Specifically the prime lattice basis is defined as 
\begin{equation}
    B_{m,c} =
    \begin{bmatrix}
    f(1) & 0 & 0 & 0 \\
    0 & f(2) & 0 & 0 \\
    \vdots & \vdots & \ddots & \vdots \\
    0 & 0 & \cdots & f(m) \\
    \lfloor 10^c \ln(p_1) \rceil &
    \lfloor 10^c \ln(p_2) \rceil &
    \cdots &
    \lfloor 10^c \ln(p_m) \rceil
    \end{bmatrix}
\end{equation}

Where $m$ is the dimension of the lattice, $c$ is the \textit{lattice-precision} parameter, $f:[1,n]\rightarrow[1,n]$ is a random permutation of $\{ \lceil 1/2 \rceil, \lceil 2/2 \rceil,\ldots, \lceil n/2 \rceil\}$, and $p_i$ is the $i$-th prime. The associated target vector $\mathbf{t}\in \mathbb{R}^m$ is defined as 

\begin{equation}
	\mathbf{t}=[0,\ldots,0, \lfloor10^c\ln(N)\rceil]
\end{equation}

Given a lattice point $\mathbf{x}=\sum_{i=1}^m e_i\mathbf{b}_i$ in a prime lattice $\mathcal{L}(B)$, where $\mathbf{b}_i$ is a basis vector of $B$, and $\mathbf{e}\in\mathbb{Z}^m$ a pair of $p_m$-smooth numbers $(u,v)$ may be obtained by

\begin{equation} \label{eq:lattice-point-to-uv}
    u=\prod_{e_i\geq0, \ i\in[1,m]} p_i^{e_i} \ ,\quad v=\prod_{e_i<0, \ i\in[1,m]} p_i^{-e_i} 
\end{equation}

The coefficient vector $\mathbf{e}$ is obtained using the pseudoinverse of the rectangular basis matrix \cite{tesoroQuantumInspiredFactorization2024c}
\begin{equation}
    \mathbf{e} = B^{-1}\cdot \mathbf{x}
\end{equation}

% ---
% Precisely, the optimization problem we seek to solve and the focus of this paper is:
% Given a prime lattice basis $B$, an target point $t$, maximize the size of the set of lattice points $S \subset \mathcal{L}(B)$ such that each lattice point $x\in S$ satisfies $\forall y \in \mathcal{L}(B), ||x-t||\leq\gamma ||y-t||$. 

% That is, find the largest set of solutions to CVP$_\gamma$. The lattice points in this set are expected to have a higher likelihood of yielding sr-pairs which can then be collected and processed in the broader factoring algorithm.

\subsection{Prime lattice parameters}
\begin{itemize}
\item \textbf{Lattice dimension,} $m$.
    Directly, $m$ specifies the dimension of the lattice. Additionally, due to the construction of the lattice, $m$ specifies the size of the factoring basis for $u,v$.

    In the original method proposed by Schnorr \cite{schnorrFactoringIntegersComputing1991, schnorrFastFactoringIntegers2021} and in a recent work by Yan et al. \cite{yanFactoringIntegersSublinear2022a}, it is claimed that it is sufficient for $m$ to grow sublinearly in the bit length of the semiprime $N$. However, this claim has been refuted by several works including \cite{khattarCommentFactoringIntegers2023, aboumradQuantumClassicalCombinatorial2023} which show that in the sublinear regime the number of sr-pairs available to be found in each lattice vanishes exponentially, therefore causing the required number of CVP solutions to increase exponentially. Detailed discussion of this point can be found in \cite{aboumradQuantumClassicalCombinatorial2023}. 

\item \textbf{Smoothness bound,} $M$.
    The size of the factoring basis used to check $u-vN$ for smoothness. A lattice point in a prime lattice corresponds to an sr-pair if $u,v,u-vN$ are all $p_M$-smooth. Furthermore, $M$ specifies how many sr-pairs must be found during the collection phase of the algorithm. $M+2$ sr-pairs are required to guarantee a congruence of squares can be generated in the linear algebra phase of the algorithm (although this number does not guarantee that congruence is nontrivial).
    
    Setting $M>m$ increases the proportion of lattice points that are sr-pairs without modifying the lattice. By the fundamental theorem of arithmetic, every value $|u-vN|$ has a prime factorization and is therefore smooth for some $M$. However, we cannot simply set a very large $M$ due to the tradeoff between the smoothness bound and the number of sr-pairs required to be found, which rises linearly in $M$.

\item \textbf{Lattice precision parameter,} $c$.
    Varying this term increases the number of prime lattices that it is possible to generate for a specific combination of $N$ and $m$. In \cite{schnorrFactoringIntegersComputing1991} it is stated that $1<c$ but without further guidance. Subsequent work has chosen to use a constant $c=4$ \cite{khattarCommentFactoringIntegers2023, priestleyPracticallyScalableApproach2025}. We have chosen to do the same.

\end{itemize}

\subsection{Optimization problems}
There are two closely related optimization problems that we seek to solve in this paper. The first is $\gamma$-CVP as presented in Definition \ref{def:gap-cvp}. As the algorithm we present is a heuristic algorithm, we do not set a specific $\gamma$ or offer a bound on $\gamma$, instead the goal is to find the best approximation within a specified amount of computation.

The second related problem is that of sr-pair collection. Precisely, given a prime lattice basis $B$, a target point $\mathbf{t}$, and a smoothness bound $M>m$, maximize the size of the set of lattice points $S \subset \mathcal{L}(B)$ such that each lattice point $\mathbf{x}\in S$ yields an sr-pair according to Equation \eqref{eq:lattice-point-to-uv}.

\section{Methodology}
% In this work we look to apply probabilistic computing methods to the optimization problems described in section \ref{sec:problem_formulation}.
We extend the method presented by Yan et al. \cite{yanFactoringIntegersSublinear2022a} in which an approximate closest vector is found via Babai's nearest plane algorithm \cite{babaiLovaszLatticeReduction1986} and further refined through a heuristic neighborhood search. In their approach, Yan et al. \cite{yanFactoringIntegersSublinear2022a} use QAOA to perform the heuristic search. The claim of quantum speed up for this task is not clear, with some critical discussions found in the literature \cite{khattarCommentFactoringIntegers2023, aboumradQuantumClassicalCombinatorial2023}. While there does not appear to be a reason to expect a quantum speedup for this task, the addition of a heuristic refinement step is a promising approach for improving the lattice-based factoring algorithm. 

In this section we present our method for applying probabilistic computing to the task of refining CVP approximations via neighborhood search. Furthermore, in Section \ref{subsec:method:lattice_parameters} we explain our selection of lattice parameters.

% First we present the general method for solving CVP and 
\subsection{Initial approximation}\label{subsec:method:solving_cvp}
\begin{algorithm}[h]
\caption{Babai's Nearest Plane Algorithm \cite{babaiLovaszLatticeReduction1986, stephens-davidowitzCVPBabaisAlgorithm2016}}\label{alg:babais}
\begin{algorithmic}[1]
    \State \textbf{Input:} target vector $\mathbf{t}$, basis vectors $\mathbf{b}_1, \ldots, \mathbf{b}_n$, and basis vector Gram-Schmidt orthogonalizations $\tilde{\mathbf{b}}_1, \ldots, \tilde{\mathbf{b}}_n$
    \State \textbf{Output:} Approximate CVP solution $\mathbf{b}_{op}$
    \State Set $\mathbf{b}_{op} \gets -\mathbf{t}$
    \For{$i = 1, \ldots, n$}
      \State $\mu \gets \langle \mathbf{b}_{op}, \tilde{\mathbf{b}}_{n-i+1} \rangle / \lVert \tilde{\mathbf{b}}_{n-i+1} \rVert^2$\label{line:gram-schmidt-coeff}
      \State $c\gets \lfloor \mu\rceil$\label{line:round}
      \State $\mathbf{b}_{op} \gets \mathbf{b}_{op} - c \mathbf{b}_{n-i+1}$ \label{line:update-approx}
    \EndFor
    \State \textbf{Output} $\mathbf{b}_{op}$
\end{algorithmic}
\end{algorithm}

Initially, the problem of finding a close vectors in a prime lattice is approximated in the standard way of approximating the CVP. First, the prime lattice basis $B$ is reduced to the lattice basis $D$ via the LLL lattice basis reduction algorithm \cite{lenstraFactoringPolynomialsRational1982}. This process converts the arbitrary structure of the basis vectors of $B$ into short, nearly orthogonal basis vectors, generating $D$. Using $D$ as the lattice basis matrix, Babai's nearest plane algorithm is applied to find a bounded approximation of $\mathbf{t}$. The approximation $\mathbf{b}_{op}$ can be efficiently computed for the approximation ratio $\gamma=2(\frac{2}{\sqrt{3}})^n$ \cite{babaiLovaszLatticeReduction1986}.

\subsection{Refining CVP approximations}\label{subsec:method:collecting_sr_pairs}
\begin{figure}[h]
    \centering
    \includegraphics[width=0.75\linewidth]{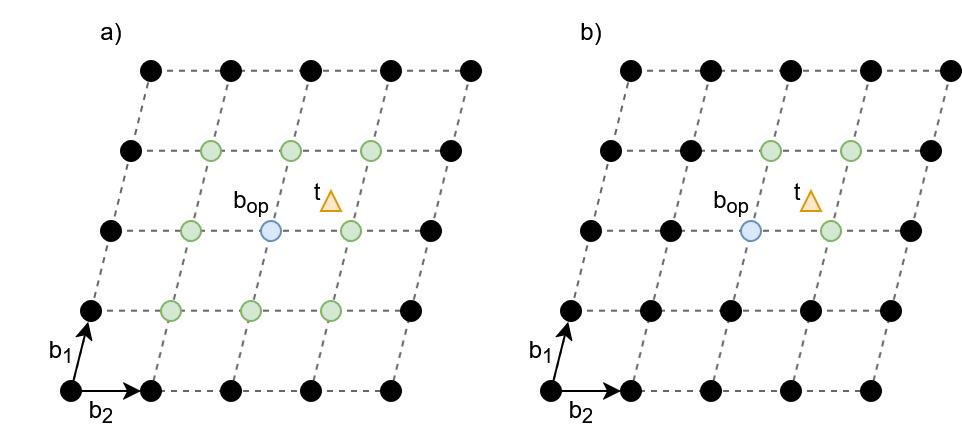}
    \caption{(a) The neighborhood (green) of the CVP approximation (blue) in a 2-dimensional lattice. Each neighbor can be reached by adding or subtracting each basis vector at most once. (b) The reduced neighborhood (green) of the CVP approximation (blue). Each basis vector is assigned a direction (both positive in this example) and each neighbor can be reached by adding or subtracting each basis vector in its assigned direction at most once.}
    \label{fig:lattice}
\end{figure}

The refinement process, first presented by Yan et al. \cite{yanFactoringIntegersSublinear2022a}, attempts to take the CVP approximation $\mathbf{b}_{op}$ and find a neighboring lattice point that is closer than $\mathbf{b}_{op}$ to $\mathbf{t}$. In this context the neighborhood of a lattice point is used to mean any lattice point that can be reached by the addition or subtraction of each LLL-reduced basis vector at most once (Figure \ref{fig:lattice}(a)). However, in order to reduce the refinement search space, the contribution of each basis vector is only considered in one direction, i.e. either added or subtracted (Figure \ref{fig:lattice}(b)). The direction of contribution of each vector is determined during the computation of Babai's nearest plane algorithm. The algorithm processes each basis vector sequentially, first calculating the Gram-Schmidt coefficient (Algorithm \ref{alg:babais}, Line \ref{line:gram-schmidt-coeff}), rounding the coefficient to the nearest integer, $c$ (Algorithm \ref{alg:babais}, Line \ref{line:round}), and updating the approximation (Algorithm \ref{alg:babais}, Line \ref{line:update-approx}). For each iteration $i$, the values $\mu_i$ and $c_i$ are recorded. The value $k_i=\operatorname{sign}(\mu_i-c_i)$ is the opposite direction to the rounding that was applied to each basis vector during Babais nearest plane algorithm (e.g. $\lceil 1.5\rfloor = 2 \implies k_i = -1$). The goal of the neighborhood search is to trial both rounding directions for each Gram-Schmidt coefficient. This is achieved by searching the set of lattice vectors
\begin{equation}
    \left\{ \mathbf{b}_{op} + \sum_{i=1}^{n} z_ik_i\mathbf{b}_i \ \middle| \ (z_1,\ldots, z_n)\in\{0,1\}^n\right\}
\end{equation}

% TODO: Include neighbourhood figure.

\subsection{Application of probabilistic computing}\label{subsec:method:probabilistic_computing}
\begin{algorithm}[h]
\caption{Calculate P-Bit Bias} \label{alg:bias}
\begin{algorithmic}[1] % The number tells where line numbering should start
    \Function{calculate\_bias}{$\mathbf{s}$, $i$, $\beta$}
    % \State $\mathbf{s}[i]$ $\gets 0$
    \State $\mathbf{v_0} \gets \mathbf{t} -\mathbf{b}_{op} - \sum_{j=1, j\neq i}^n s_jk_j\mathbf{b}_j$ 
    \State $\mathbf{v_1}$ $\gets \mathbf{v_0} + k_i\mathbf{b}_i$ \label{line:calc_bias_2} 
    \State $e_0 \gets \mathbf{v_0}\cdot\mathbf{v_0}$ \label{line:calc_bias_3}
    \State $e_1 \gets \mathbf{v_1}\cdot\mathbf{v_1}$ \label{line:calc_bias_4}
    \State \Return $\beta \cdot(e_0-e_1)$ \label{line:calc_bias_5}
  
    \EndFunction
\end{algorithmic}
\end{algorithm}

\begin{figure}[h]
    \centering
    \includegraphics[width=0.5\linewidth]{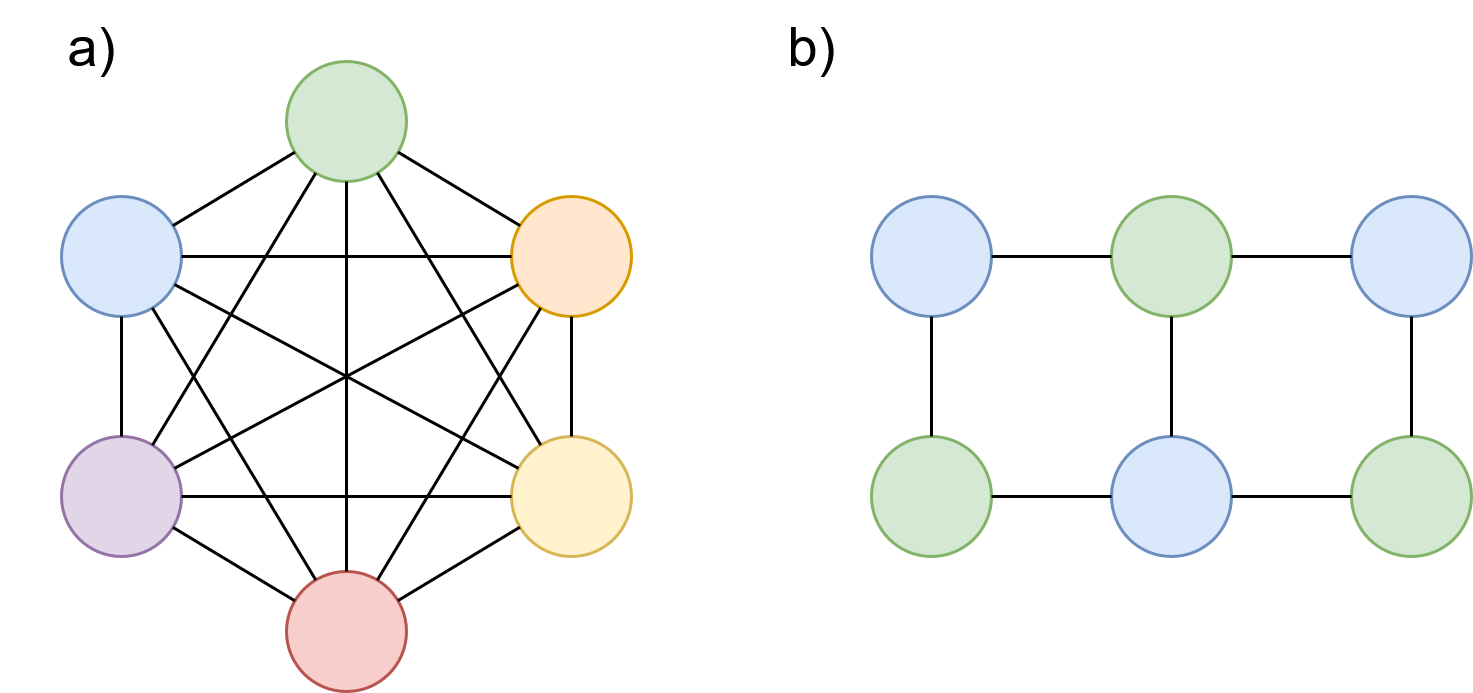}
    \caption{(a) A fully connected p-bit network. The bias calculation of each p-bit is dependent on the state of every other p-bit in the network. (b) A 2D lattice p-bit network. The bias calculation of each p-bit is dependent on only the neighbors of the given p-bit. In this configuration, non-neighboring p-bits can safely update in parallel. In a general p-bit network, graph coloring is used to identify sets of parallelizable update groups.}
    \label{fig:pbit_neworks}
\end{figure}

Probabilistic computing is a model of computation wherein the basic unit of information is the p-bit, as per Definition \ref{def:p-bit}. The state of the p-bit randomly flips between state $0$ and state $1$ according to a specified bias value. A collection of p-bits can be networked together in a graph structure such that the neighbors of a given p-bit influence the bias of said p-bit. The bias of a p-bit is calculated according to Definition \ref{def:pbit-bias}, such that the bias is the difference in the energy of the system when the given p-bit is in the $0$ state and in the $1$ state.

The system evolves by repeatedly selecting a p-bit according to a sampling algorithm and calculating a new bias from the instantaneous state of the neighborhood of the selected p-bit. As each bias represents the difference in system energy contributed by a given p-bit, the system iterates towards a set of p-bit biases that yield system states with low energies. In this way, probabilistic computing can perform heuristic combinatorial optimization in a similar capacity to algorithms such as simulated annealing, quantum annealing, or QAOA.

In order to apply probabilistic computing to CVP approximation refinement, we first define the energy equation and derive the bias calculation. The objective in the CVP is to find $\mathbf{x} \in \mathcal{L}$ such that the distance between $\mathbf{x}$ and $\mathbf{t}$ is minimized (Definition \ref{def:gap-cvp}). Therefore, we use the euclidean norm as the basis for the system energy. Specifically, as it is an order preserving operation on positive numbers, we use the square of the euclidean norm to avoid computing the square root \cite{yanFactoringIntegersSublinear2022a}. 

\begin{equation} \label{eq:vcbm-energy-func}
    E(\mathbf{s}) =\left|\left| \mathbf{t}-\left(\mathbf{b}_{op} + \sum_{i=1}^{n} s_ik_i\mathbf{b}_i\right)\right|\right|^2 
\end{equation}

Where $\mathbf{s}$ is the global state of the system, i.e. the sampled instantaneous state of each p-bit according to its current bias. Furthermore, pseudocode for the bias calculation is shown in Algorithm \ref{alg:bias}. The value $\beta$ is included in Algorithm \ref{alg:bias} Line \ref{line:calc_bias_5} to represent the inverse temperature, as used in many combinatorial solvers \cite{vanlaarhovenSimulatedAnnealingTheory1987}.

Note that from the bias calculation in Algorithm \ref{alg:bias} we can see that the neighborhood of each p-bit is every other p-bit in the system as the bias calculation for p-bit $i$ depends on $s_j$ for $j=1,\ldots, i-1,i+1, \ldots n$. The p-bit network is therefore a fully connected graph (Figure \ref{fig:pbit_neworks}). This is a limitation of this method, as one potential advantage of probabilistic computing is the capability to perform parallel bias updates for non-neighboring p-bits \cite{hasselgrenProbabilisticComputingOptimization, searleVirtuallyConnectedProbabilistic}.

The full algorithm for CVP approximation refinement via probabilistic computing can be found in Appendix \ref{appendix:algorithm}.

\subsection{Discussion of lattice parameters} \label{subsec:method:lattice_parameters}
The lattice dimension $m$ and the smoothness bound $M$ are two interrelated parameters for constructing prime lattices during the collection phase of lattice-based factoring. In Schnorr's original algorithm $m$ and $M$ are equivalent and scale sublinearly in the bit length of the semiprime $N$. However, it has been shown that this requires solving exponentially many lattices due to the fact the smoothness bound does not grow fast enough  and the number of available sr-pairs in a lattice disappears exponentially \cite{aboumradQuantumClassicalCombinatorial2023}. 

More recent work suggests separating $m$ and $M$ such that $m<M$ \cite{yanFactoringIntegersSublinear2022a}. This increases the number of available sr-pairs per lattice as it increases the smoothness bound at the cost of requiring more sr-pairs to be found. With an appropriate choice of $M$ this may reduce the overall computation, but in \cite{yanFactoringIntegersSublinear2022a} only three concrete values of $M$ are provided without justification or method for generation. Further work has suggested using $M=m^2$ \cite{aboumradQuantumClassicalCombinatorial2023}.

One noted issue with this approach is the number of sr-pair collisions that occur, i.e. the number of times the same sr-pair is found in two distinct lattices. The number of available pairs $u,v$ increases proportionally to $m$ as by construction each lattice point corresponds to a distinct pair $u,v$. Therefore, when $m$ increases at a reduced rate compared to $M$, we expect more sr-pair collisions to occur as fewer unique triplets $u,v,u-vN$ are available to be considered.

The choice of $m$ and $M$ is informed by the hardware constraints of the heuristic solver used for the refinement process. In \cite{yanFactoringIntegersSublinear2022a} the primary objective was to reduce the number of qubits needed. Ideally, these parameters would evolve such that the lattice dimension grows at a rate fast enough to minimize collisions and the smoothness bound is selected to be an optimal trade off between the total number of required sr-pairs ($M+2$) and the expected number of sr-pairs per lattice.

% TODO: state selected params

% maximized to avoid the number of available sr-pairs does decay exponentially as in the sublinear scheme.

\section{Experiments and Results}

The objectives of our experiments are twofold. First, we aim to determine suitable parameters for generating prime lattices. Second, we aim to assess the suitability of probabilistic computing for CVP approximation refinement and furthermore to the application of sr-pair collection. Our experiments were conducted using a software simulation implementation of probabilistic computing.  

Due to the intractability of the CVP, data sets were generated at a size that could be completed within several days given the available hardware. Unless otherwise specified, we anticipate that increasing the data set size would primarily lead to a reduction in the width of the confidence interval. Error bars represent a 95\% confidence interval. All data sets used are available at \cite{al-hassoProbabilisticComputingApproach2025}.

\subsection{Lattice dimension and smoothness bounds}\label{subsec:exp-lattice-dim-and-smoothness-bounds}
\begin{figure}[t]
    \centering
    \includegraphics[width=1\linewidth]{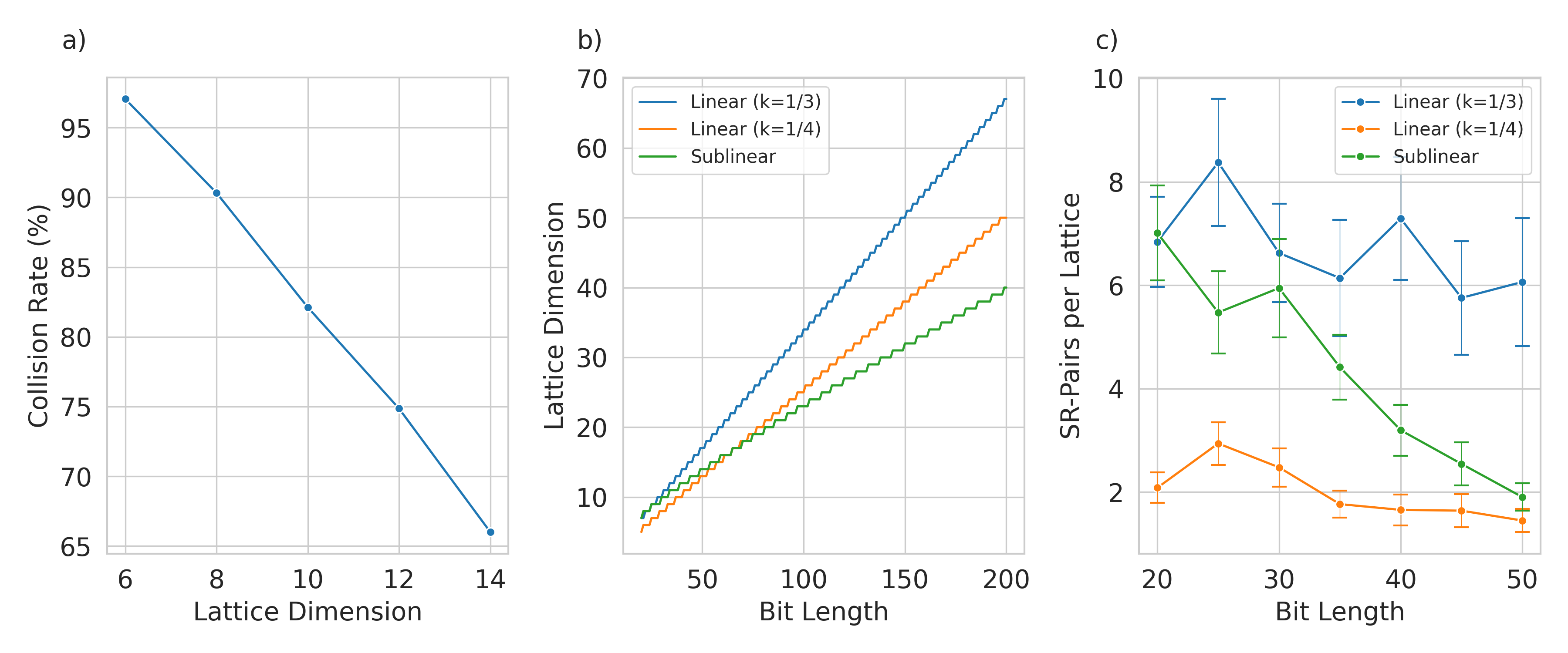}
    % \caption{(a) The collision rate of sr-pairs across 500 lattice instances against lattice dimension. (b) Bit length to dimension mappings considered in this paper. Linear mapping $m=\lceil k\cdot\text{bit length}\rceil$. Sublinear mapping: $m=\lceil\frac{3}{2}\cdot\frac{\text{bit length}}{\log_2(\text{bit length})}\rceil$. (c) The mean number of sr-pairs found per lattice across 100 lattices per bit length and mapping.}
    \caption{\textbf{(a)} Sr-pair collision rate for a fixed semiprime $N$ with bit length 30 for 500 lattices per data point. \textbf{(b)} Bit length to dimension mappings considered in this work. Linear mapping: $m=\lceil k\cdot\text{bit length}\rceil$. Sublinear mapping: $m=\lceil\frac{3}{2}\cdot(\text{bit length}/\log_2(\text{bit length}))\rceil$. \textbf{(c)} Mean number of sr-pairs per lattice for 100 lattices per data point.}
    \label{fig:exp2}
\end{figure}
% one key issue is collisions, a larger lattice leads to fewer collisions (but all lattice points are smooth in some factoring basis)

% Two of the key hyperparameters for an instance of prime lattice CVP is the lattice dimension, $m$ and the smoothness bound $M$. The values $u,v$ associated with a lattice point are $p_m$-smooth by construction but by using $M>m$ more triplets $u,v,u-vN$ can be considered sr-pairs as there are a greater number of values for $u-vN$ that are $p_M$-smooth. By the fundamental theorem of arithmetic, every value $|u-vN|$ is a product of prime numbers and is therefore smooth for some smoothness bound $M$.

% Previous work on sr-pair collection in prime lattices has focused on scaling the lattice dimension $m$ sublinearly in the bit length of the semiprime $N$ with the smoothness bound $M > m$ (CITE). In Schnorr's algorithm, $m=M$ and the choice to set $M>m$ comes from a hardware constant of quantum computers and a need to minimize the number of qubits used in the algorithm.

% Sublinear scaling was chosen so that the number of required lattices would grow sub-exponentially, but recent results have shown this to be 

We performed two experiments to support the discussion in Section \ref{subsec:method:lattice_parameters}. In the first experiment, we consider a semiprime $N$ with bit length 30. The smoothness bound was consistent across dimensions, set at $M=14^2$ as $14$ was the largest dimension tested. By enumeration, each candidate lattice point available in the search space of each lattice was then tested for smoothness. The results of this experiment are shown in Figure \ref{fig:exp2}(a), showing a negative linear relationship between the collision rate and lattice dimension. Furthermore, the collision rate is a function of the size of the data set, as there are a finite number of unique sr-pairs to be found after which every additional pair is necessarily a collision. The data set size of 500 lattices was chosen arbitrarily.

In the second experiment, we compare the number of sr-pairs available per lattice in different bit-length-to-lattice-dimension mappings. We compare the sublinear mapping of \cite{yanFactoringIntegersSublinear2022a} with two linear mappings, shown in Figure \ref{fig:exp2}(b). The sr-pairs in each lattice were found by enumeration, using a smoothness bound of $M=m^2$. The results of this experiment are shown in Figure \ref{fig:exp2}(c). All three schemes show a decrease in the mean number of sr-pairs, with the sublinear scheme having the highest rate of decrease. Comparing Figure \ref{fig:exp2}(b) and Figure \ref{fig:exp2}(c) we can see that the average number of sr-pairs per lattice corresponds to the lattice dimension yielded by each of the three mappings. For the remaining experiments we disregard the sublinear mapping.
% In an ideal system the smoothness bound should increase at a rate such that the number of sr-pairs does not disappear exponentially and the difference $M-m$ should be minimized to reduce the effect of collisions.

\subsection{Probabilistic computing for CVP approximation refinement}

\begin{figure}[t]
    \centering
    \includegraphics[width=1\linewidth]{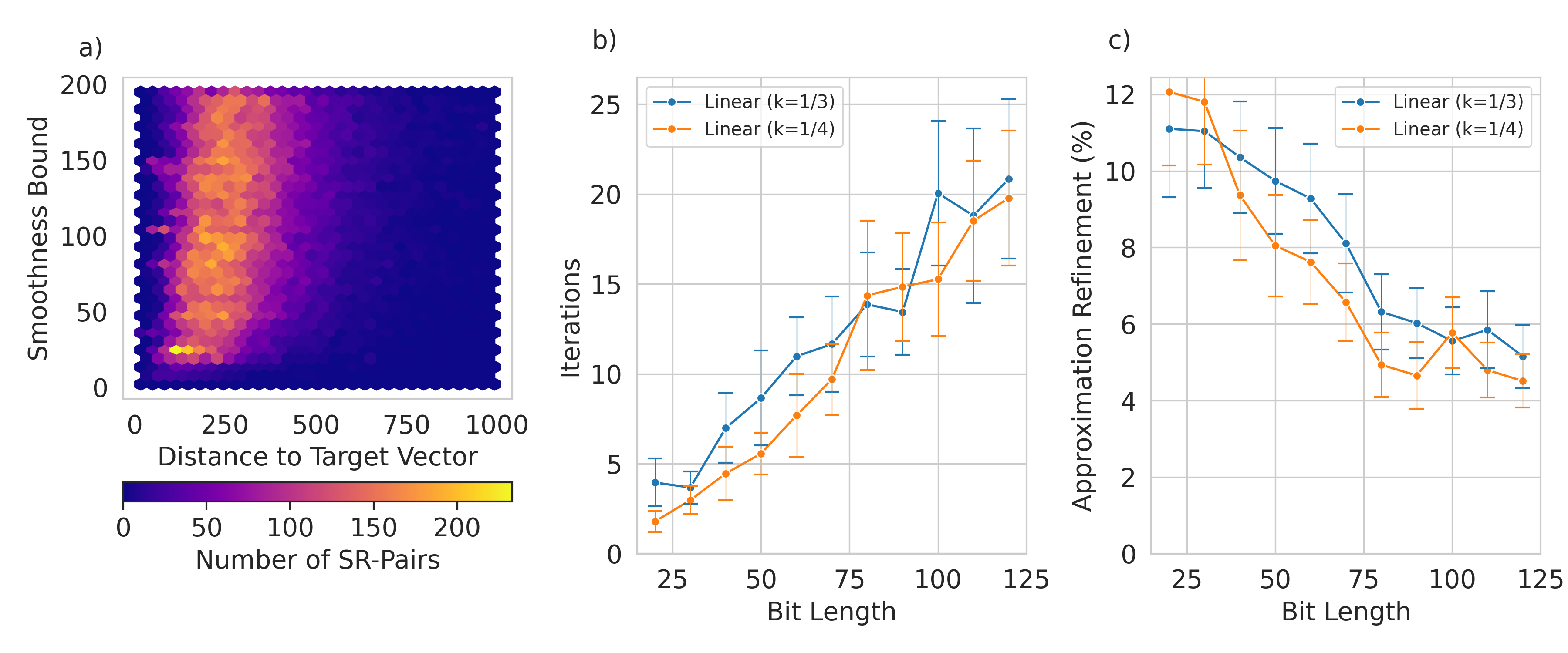}
    % \caption{(a) The average number of iterations required to find maximal refinement in a lattice across 100 lattices per mapping and bit length. (b) The average percentage improvement in distance to target of the initial CVP approximation found by the probabilistic computer across 100 lattices per mapping and bit length. (c) Density of sr-pairs across 500 lattices of varying dimension, as found by enumeration.}
    \caption{\textbf{(a)} Number of sr-pairs as a function of distance to target vector and smoothness bound for 500 lattices of varying dimensions. \textbf{(b)} Mean iterations required to find maximal CVP approximation refinement for 100 lattices per data point. \textbf{(c)} Mean percentage improvement in distance to target over initial approximation for refinements found in (b). }
    \label{fig:exp3}
\end{figure}
% We next determine the suitability of using probabilistic computing as a heuristic optimizer for refining CVP. To measure the performance of the probabilistic algorithm, we generated a data set of X lattice instances for each bit length where the lattice dimension scales linearly in the bit length, $m=\frac{1}{3}\cdot\log_2(N)$. Only lattices with a refinement to be found in the neighborhood of Babai's approximation were considered for this experiment. Where it was not possible to check for the existence of a refinement via enumeration due to lattice dimension, lattice points that are the sum of babai's approximation and fewer than $6$ basis vectors were considered. This vastly reduces the required search space and allowed us to test the VCBM on lattices up to dimension X.

We next determine the suitability of using probabilistic computing as a heuristic optimizer for refining CVP approximations. To measure the performance of the probabilistic computing algorithm, we generated a set of lattices that were found by enumeration to contain at least one possible refinement. We tested both linear mappings.

We are interested in finding close vectors as the lattice-based factoring algorithm is predicated on the claim that the distance between a lattice point and the target is correlated with the likelihood of that lattice point being an sr-pair due to the lattice construction \cite{schnorrFactoringIntegersComputing1991, schnorrFastFactoringIntegers2021}. Figure \ref{fig:exp3}(a) supports this claim showing that the number of sr-pairs in a lattice increases as the distance to the target vector decreases, until reaching a threshold.

In this experiment the probabilistic computer was configured to execute with a linearly increasing $\beta$ value (Algorithm \ref{alg:bias} Line 6). Analogous to simulated annealing \cite{vanlaarhovenSimulatedAnnealingTheory1987}, $\beta$ controls the likelihood of the system moving to higher energy states. As $\beta$ increases, the system converges towards a minimum. For every lattice tested, the probabilistic computer found the maximal refinement as determined by enumeration\footnote{We observed that, in the vast majority of cases, a refinement was found due to the contribution of $6$ or fewer basis vectors. Therefore, when enumerating large lattices we checked only the states with $6$ or fewer p-bits in the $1$ state.}. Figure \ref{fig:exp3}(b) shows the number of iterations required to find the maximal refinement, where an iteration consists of a full-sweep of p-bit updates. Furthermore, the percentage improvement in distance to target, i.e. quality of CVP approximation refinement, is shown in figure \ref{fig:exp3}(c). These results show that the percentage improvement in the CVP approximation decreases as the bit length (i.e. lattice dimension) grows. From the bit lengths tested, we have not found whether the percentage refinement converges.

% The cost of one iteration increases like $\Theta(m^2)$ as both the number of p-bits to update per sweep and the cost of a p-bit update increase linearly in $m$. As the number of iterations required to find the maximal refinement appears to increase linearly, we determine the expected time to solution would increase like $\Theta(m^3)$. 

The number of p-bits in the system increases linearly in $m$, meaning both the number of p-bits to update per iteration and the cost of performing one bias calculation (Algorithm \ref{alg:bias}) scale like $\Theta(m)$. Thus, with the results from Figure \ref{fig:exp3}(b) indicating a plausible linear scaling of iterations required for attaining maximal refinement, we expect the probabilistic computing architecture to attain a time to solution scaling like $\Theta(m^3)$.

\subsection{Probabilistic computing for sr-pair collection}

\begin{figure}[t]
    \centering
    \includegraphics[width=1\linewidth]{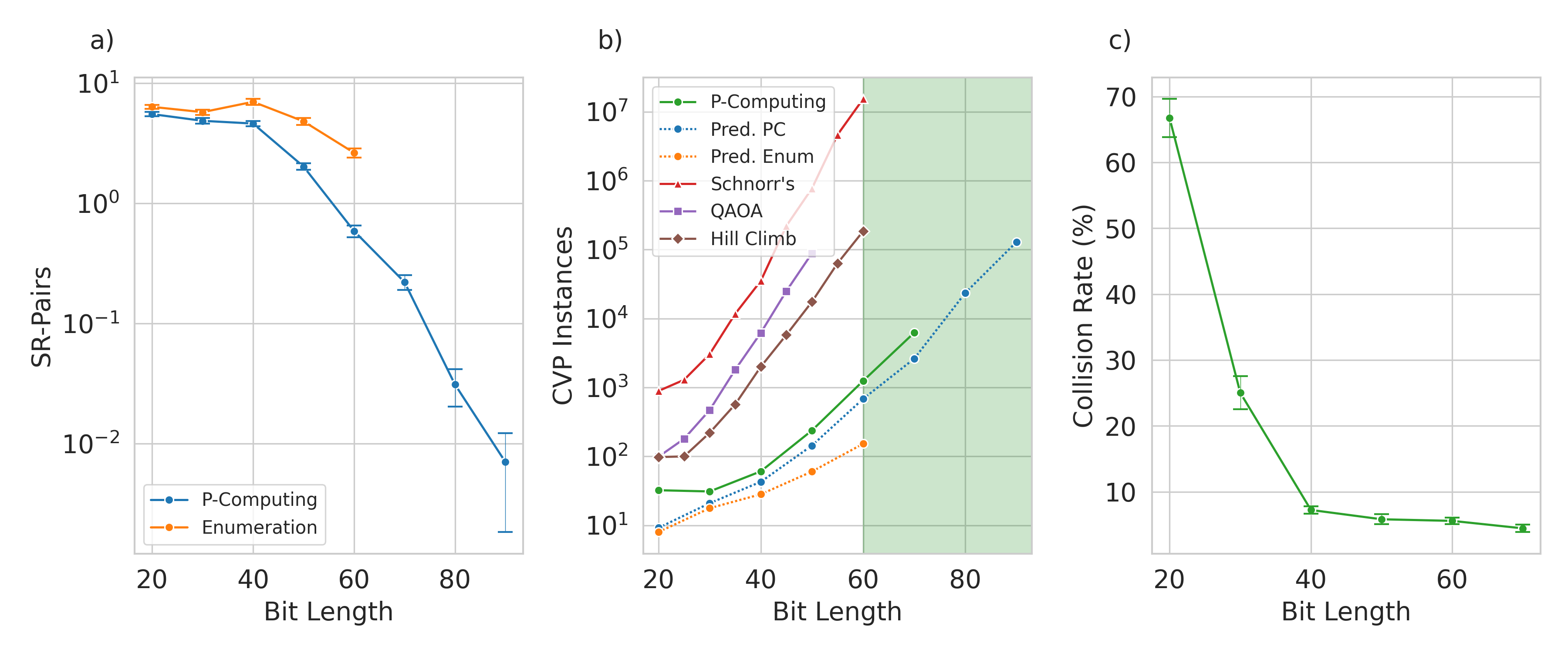}
    % \caption{(a) The average number of sr-pairs found by the probabilistic computer compared to the total number of possible sr-pairs determined by brute force where feasible. (b) The expected number of lattices required to find $M+2$ sr-pairs, and thus generate a congruence of squares, using probabilistic computing compared against the number of lattice instances required by the three heuristic solvers reported in \cite{aboumradQuantumClassicalCombinatorial2023} (Schnorr's, QAOA, and Hill Climb) and a hypothetical perfect solver capable of finding every sr-pair in a lattice.}
    \caption{\textbf{(a)} Mean sr-pairs per CVP instance found by probabilistic computing and by enumeration where feasible for 1,000 lattices per data point. \textbf{(b)} Mean number of CVP instances required to factor a semiprime using various heuristic refinement methods: Probabilistic Computing (this work), Predicted Probabilistic Computing (this work), Predicted Enumeration (this work), Schnorr's algorithm \cite{schnorrFactoringIntegersComputing1991, schnorrFastFactoringIntegers2021}, Yan et al.'s QAOA method \cite{yanFactoringIntegersSublinear2022a}, and a classical hill-climbing algorithm. For probabilistic computing, 25 semiprime integers were factored per data point. Results for Schnorr's algorithm, QAOA, and hill climbing are reported from \cite{aboumradQuantumClassicalCombinatorial2023}. \textbf{(c)} Mean collision rate of sr-pairs for semiprime factoring via probabilistic computing for the same instances shown in (b).}
    \label{fig:exp4}
\end{figure}

% To apply the probabilistic computer to the task of collecting SR-pairs, for each lattice point considered we calculate the values $u,v$ and $u-vN$. 

Several modifications were made to the CVP approximation refinement algorithm in order to apply the probabilistic computer to the task of collecting sr-pairs. First, after each bias update, the current global state of the probabilistic computer is recorded. From this, the corresponding lattice point is derived and the values $u, v, u-vN$ are determined. The value $u-vN$ is then checked for smoothness up to the smoothness bound $M$. This computational load is non-blocking, therefore the probabilistic computer can continue its search operation while candidate states are processed in parallel. We note that lattice points closer to the target vector are expected to have a higher likelihood of yielding sr-pairs (Figure \ref{fig:exp3}(a)). Therefore, should checking candidate lattice points become a computational bottleneck, candidates can be processed in ascending cost order to prioritize higher likelihood candidates. 

% Furthermore, the candidate states can be processed in ascending cost order as we expect lattice points close to the target vector to have a higher likelihood of being an sr-pair. %TODO include hexplot figure.

Two further modifications were made, to the value of the inverse temperature $\beta$ and the stopping criterion of the algorithm. In the CVP approximation refinement algorithm, the objective is to simply find the closest lattice point to the target vector $t$. In the above experiment $\beta$ increased linearly and the system terminated when the lattice point corresponding to the maximal refinement was found, as precomputed by enumeration. In the sr-pair search version of the algorithm, while the bias calculation remains the same, the goal is instead to find the maximum number of lattice points that are sr-pairs. This requires us to search many, ideally close, candidate lattice points rather than converging on a global minima. We chose to use a fixed value of $\beta=0.66$, and a number of iterations that increases linearly according to $iterations = 20m$. The number of bias updates therefore increases like $\Theta(m^2)$ as both the number of full sweeps and the number of p-bits increase linearly in bit length. We continue to use a linear bit length to dimension mapping with $k=\frac{1}{3}$ and choose to use $M=m^2$ as suggested in \cite{aboumradQuantumClassicalCombinatorial2023}. Our choice of parameters was guided by preliminary experiments, but we make no claims of optimality in our choice.

%Here we use iterations to mean the number of `full sweeps' where the bias of each p-bit is updated once. 

%dimension scaling of $m=\frac{1}{3}\cdot\log_2(N)$

% For this experiment, 1,000 lattice instances were tested per bit length.
Figure \ref{fig:exp4}(a) shows the mean number of sr-pairs found per lattice at each bit-length. For bit lengths 20 to 60, the lattices were also enumerated to determine the true number of sr-pairs available per lattice. On average, the probabilistic computer found $66.9\%$ of available sr-pairs as found by enumeration. However, despite increasing the rate of growth of lattice dimension from sublinear to linear, the number of available sr-pairs and the number of sr-pairs found by this method vanishes exponentially. In the full context of lattice-based factoring this implies that the number of lattices that need to be tested would increase exponentially.
% as shown in Figure \ref{fig:exp4}(b). 

Figure \ref{fig:exp4}(b) shows the number of instance of CVP required to factor a semiprime number using different methods. Using the data from Figure \ref{fig:exp4}(a), we make two predictions for how many CVP instances are required to find $M+2$ sr-pairs. The first (Pred. Enum) for an intractable perfect solver that finds every sr-pair in a lattice by enumeration. This is provided as a reference series. The second (Pred. PC) is for the probabilistic computer, based on the average number of sr-pairs found per lattice in the previous experiment. Both of these predictions ignore the effect of sr-pair collisions. Our main result is the mean number of lattice instances required by the probabilistic computer (P-Computing) to factor semiprime numbers. At bit length 60, this probabilistic computing method requires two orders of magnitude fewer CVP instances than the best method reported in \cite{aboumradQuantumClassicalCombinatorial2023}. 

Figure \ref{fig:exp4}(c) shows the mean sr-pair collision rate of the probabilistic computer during semiprime factoring. For small bit lengths, i.e. small lattice dimensions, the collision rate is large. As the bit length grows larger, the collision rate drops to $\sim5\%$. This is in line with our discussion in Section \ref{subsec:method:lattice_parameters} and results in Section \ref{subsec:exp-lattice-dim-and-smoothness-bounds} wherein lattices with reduced dimension exhibit higher collision rates. In Figure \ref{fig:exp4}(b), the difference between the actual performance of the probabilistic computer and the predicted performance is due to the collision rate. As such, beyond bit length 70, we expect (P-Computing) to continue to scale like (Pred. PC).

% Even though the number of lattices required does increase exponentially, the number of lattices required is several orders of magnitude smaller and grows a reduced rate when compared to the approach presented in \cite{yanFactoringIntegersSublinear2022a}, according to the analysis of \cite{aboumradQuantumClassicalCombinatorial2023}.

% \subsubsection{Hardware}
% TODO

% \section{Discussion}
% TODO: Hardw

\section{Conclusion and Outlook}

In this work we investigated the application of probabilistic computing to solving the closest vector problem in the context of lattice-based factoring. We presented a discussion on the selection of parameters for this method, supported by experimental data. Furthermore, we found that probabilistic computing is well suited to the task of heuristic CVP approximation refinement and to the task of sr-pair collection. Extending the lattice based factoring method presented in \cite{schnorrFactoringIntegersComputing1991, schnorrFastFactoringIntegers2021} and further developed in \cite{yanFactoringIntegersSublinear2022a}, the method presented in this work displays significantly favorable scaling with respect to the number of lattice instances required to be solved during the factoring collection phase.

% Despite this improvement, it is clear that there remain several limitations to this method. Primarily, the number of lattices required to find the prime factors of a semiprime integer 

Despite the improvement that this method offers in the refinement of CVP solutions, it is clear that there remain limitations to the lattice based factoring method. Primarily, that the number of lattice required to find the prime factors of a given semiprime increases exponentially. 

% Several directions remain open for future work, including:
% \begin{itemize}
%     \item Testing and benchmarking this method on dedicated probabilistic computing hardware.
%     \item Determining the optimal scaling for lattice dimension and smoothness bound parameters with respect to minimizing collisions between lattices and maintaining a sufficient number of available sr-pairs.
%     \item Applying CVP approximation refinement using probabilistic computing to lattice-based PQC. This work focused only on the CVP using prime lattice construction, this method could be extended to general CVP instances.
%     \item Using p-bit network sparsification techniques \cite{aaditMassivelyParallelProbabilistic2022} to prevent the computational cost of the p-bit bias function increasing in the size of the system.
%     % \item Benchmarking probabilistic computing as a method of refinement against other methods such as QAOA \cite{priestleyPracticallyScalableApproach2025} and hill-climbing classical algorithms \cite{aboumradQuantumClassicalCombinatorial2023}.
% \end{itemize}

Several directions remain open for future work. These include testing and benchmarking this method on dedicated probabilistic computing hardware; determining the optimal scaling for lattice dimension and smoothness bound parameters to minimize collisions between lattices while maintaining a sufficient number of available sr-pairs; applying CVP approximation refinement using probabilistic computing to lattice-based PQC - while this work focused only on the CVP using prime lattice construction, this method could be directly applied to general CVP instances; and using p-bit network sparsification techniques \cite{aaditMassivelyParallelProbabilistic2022} to prevent the computational cost of the p-bit bias function from increasing with the size of the system.

\newpage
\printbibliography

\newpage
\appendix
\section{CVP Approximation Refinement Algorithm} \label{appendix:algorithm}

Algorithm \ref{alg:CVP-refinement} shows the full algorithm for the CVP refinement via probabilistic computing. The function READ\_PBIT is a hardware operation that can be simulated according to the process in Definition \ref{def:p-bit}. The function SMOOTHNESS\_CHECK implements the process described in Definition \ref{def:congruence-of-squares}. This function is non-blocking and can be handled in parallel. Furthermore, the candidate states can be ordered in ascending order of distance to target for processing. The function UPDATE\_BETA is an arbitrary annealing cooling schedule for the system. In this work we use both a linear decay and constant $\beta$. The function STOPPING\_CRITERION is an arbitrary exit condition for the system, typically once a predetermined number of iterations have been executed. Future systems could use significantly more sophisticated stopping criteria.

\begin{algorithm} 
\caption{CVP Approximation Refinement} \label{alg:CVP-refinement}
\begin{algorithmic}[1] % The number tells where line numbering should start
    \Require $n \in \mathbb{N}$,  $\mathbf{b} \in \mathbb{Z}^{n+1}$, $\mathbf{t}\in \mathbb{R}^{n+1}$, $\mathbf{r} \in \{-1,1\}^n$, $\mathbf{D}\in\mathbb{Z}^{(n+1)\times n}$
    \Ensure Set of sr-pairs

    \Function{select\_pbit}{$ $}
    \State $i$ $\gets$ randomly select from set of all p-bit indices
    \State \Return $i$
    \EndFunction

    \Function{update\_state}{$\mathbf{s}$, $i$}
    \State $bias \gets$ $biases[i]$
    \State $\mathbf{s}[i] \gets$ \Call{read\_pbit}{bias} 
    \State \Return $\mathbf{s}$
    \EndFunction

    \Function{update\_neighborhood\_state}{$\mathbf{s}$, $i$}
    \For{$j$ in neighborhood of $i$}
    \State $\mathbf{s}\gets $\Call{update\_state}{$j$}
    \EndFor
    \State \Return $\mathbf{s}$
    \EndFunction

    \Function{calculate\_bias}{$\mathbf{s}$, $i$, $\beta$}
    % \State $\mathbf{s}[i]$ $\gets 0$
    \State $\mathbf{v_0} \gets \mathbf{t} -\mathbf{b}_{op} - \sum_{j=1, j\neq i}^n s_jk_j\mathbf{b}_j$ 
    \State $\mathbf{v_1}$ $\gets \mathbf{v_0} + k_i\mathbf{b}_i$ 
    \State $e_0 \gets \mathbf{v_0}\cdot\mathbf{v_0}$
    \State $e_1 \gets \mathbf{v_1}\cdot\mathbf{v_1}$
    \State \Return $\beta \cdot(e_0-e_1)$ 
  
    \EndFunction
    \\

    % \Function{smoothness\_check}{$state$}
    % \EndFunction
    % \\

    \State $\mathbf{s} \gets$ array of length $n$ with elements in $\{0,1\}$, initialize to all zeros
    \State $biases \gets $ array of length $n$ with elements in $\mathbb{R}$, initialize to all zeros
    \State $\beta \gets$ a real number
    \While{\textbf{not} \Call{stopping\_criterion}{$ $}} 
    \State $i$ $\gets$ \Call{select\_pbit}{$ $}
    \State $\mathbf{s}\gets $ \Call{update\_neighborhood\_state}{$\mathbf{s}$, $i$}
    % \State $\mathbf{s} \gets$ \Call{read\_global\_state}{$ $}
    \State $biases[i]$ $\gets$ \Call{calculate\_bias}{$\mathbf{s}$, $i$, $\beta$}
    \State \Call{smoothness\_check}{$\mathbf{s}$} 
    \State $\beta \gets$ \Call{update\_beta}{$ $} 
    \EndWhile

    \State \Return
\end{algorithmic}
\end{algorithm}

\end{document}